\documentclass[notitlepage,superscriptaddress]{revtex4-1}
\usepackage{color}
\usepackage{graphics,graphicx,epsfig}
\usepackage{epsf,epstopdf,wrapfig}
\usepackage{amssymb,amsfonts,amsmath}
\usepackage[export]{adjustbox}
\usepackage{ifthen}	

\usepackage[utf8]{inputenc} 
\usepackage[T1]{fontenc}    
\usepackage{hyperref}       
\usepackage{url}            
\usepackage{booktabs}       
\usepackage{amsfonts}       
\usepackage{nicefrac}       

\usepackage{setspace}

\newcommand{\beginsupplement}{
\setcounter{table}{0}
\renewcommand{\thetable}{S\arabic{table}}%
\setcounter{figure}{0}
\renewcommand{\thefigure}{S\arabic{figure}}%
}

\begin{document}

\title{Reassessing the Exon-Foldon correspondence using Frustration Analysis}

\author{Ezequiel A. Galpern}
\affiliation{Protein Physiology Lab, Departamento de Qu\'\i mica Biol\'ogica, Facultad de Ciencias Exactas y Naturales, Universidad de Buenos Aires, C1428EGA, Buenos Aires, Argentina}
\affiliation{ Instituto de Qu\'\i mica Biol\'ogica de la Facultad de Ciencias Exactas y Naturales, CONICET - Universidad de Buenos Aires, C1428EGA, Buenos Aires, Argentina}
\author{Hana Jaafari}
\affiliation{Center for Theoretical Biological Physics, Rice University, Houston, TX 77005}
\affiliation{Applied Physics Graduate Program, Smalley-Curl Institute, Rice University, Houston, Texas 77005}

\author{Carlos Bueno}
\affiliation{Center for Theoretical Biological Physics, Rice University, Houston, TX 77005}
 
\author{Peter G. Wolynes}
\affiliation{Center for Theoretical Biological Physics, Rice University, Houston, TX 77005}
\affiliation{Department of Chemistry Rice University, Houston, TX 77005}
\affiliation{Department of Physics, Rice University, Houston, TX 77005}
\author{Diego U. Ferreiro}
\affiliation{Protein Physiology Lab, Departamento de Qu\'\i mica Biol\'ogica, Facultad de Ciencias Exactas y Naturales, Universidad de Buenos Aires, C1428EGA, Buenos Aires, Argentina}
\affiliation{ Instituto de Qu\'\i mica Biol\'ogica de la Facultad de Ciencias Exactas y Naturales, CONICET - Universidad de Buenos Aires, C1428EGA, Buenos Aires, Argentina}

\begin{abstract}
Protein folding and evolution are intimately linked phenomena. Here, we revisit the concept of exons as potential protein folding modules across 38 abundant and conserved protein families. Taking advantage of genomic exon-intron organization and extensive protein sequence data, we explore exon boundary conservation and assess their foldon-like behavior using energy landscape theoretic measurements. We found deviations in exon size distribution from exponential decay indicating selection in evolution. We describe that there is a pronounced independent foldability of segments corresponding to conserved exons, supporting the exon-foldon correspondence. We further develop a systematic partitioning of protein domains using exon boundary hot spots, unveiling minimal common exons consisting of uninterrupted alpha and/or beta elements for the majority but not all of the studied families. 
\end{abstract}


\maketitle

\section*{Significance}
If globular protein domains consist of smaller units, folding and evolution would be facilitated. The fact that natural eukaryotic proteins are genetically partitioned in exons suggests that these may correspond with foldable regions. Here we revisit the correspondence between exons and foldons, quasi independent folding units, using concepts derived from energy landscape theory. We find that conserved exons are more foldable than other partitions of the primary structure. We describe that exon boundaries rarely interrupt the continuous secondary structures in the folded domain in most but not all of the protein families analyzed.

\section*{Introduction}
Protein evolution and folding are two intertwined aspects of a complex problem. Over the past decades, a reasonable shortcut to simplify this problem has been to try to break down protein structures into distinct modules. In 1973, Wetlaufer proposed that the initial stages of folding nucleation may occur independently in separate regions \cite{wetlaufer1973nucleation}. Addressing Levinthal's paradox, he claimed that if there were individual modules that fold in parallel, the searching time for folding the entire molecule can be exponentially reduced and would be comparable to the isolated segments folding time.  With the discovery of silent DNA interrupting coding regions in Eukarya, Gilbert \cite{gilbert1978genes} and Blake \cite{blake1978genes} posited that if genes resemble a mosaic divided into pieces, then the coding pieces -christened by Gilbert ''exons''- can reasonably be expected to translate into integrally folded protein pieces, such as domains or supersecondary structures. These fragments could then shuffle and combine over evolutionary timescales, giving rise to novel functional proteins. Indeed, exons of several proteins were early characterized as structural units, including hemoglobin \cite{go1981correlation}.  Among various theories, it has been argued that exon-shuffling may have played a significant role in metazoan evolution, coinciding with a burst of evolutionary creativity during the emergence of multicellularity \cite{patthy1999genome}. 

Energy landscape theory explains how proteins fold within relevant timescales using parallel paths without explicitly dividing the molecule into parts. When a polymer is minimally frustrated, parallel search can be done in a delocalized manner as native contacts can guide the polymer folding \cite{bryngelson1987spin}. Of course, some paths may be modestly favored over others, and these variations have been successfully predicted by perfectly funneled models \cite{wolynes2005energy}. Different protein regions may fold at different times quasi-independently if the sufficiently strong native interactions largely contained within them can overcome their entropy loss. These units then may fold in a single cooperative step which have been called \emph{foldons} by Panchenko et al \cite{panchenko1996foldons}. Using a simple energy field model and a searching algorithm, they assigned foldons to many proteins and they compared them with exons. They found only a weak correlation between the evolutionary units and the folding regions \cite{panchenko1996foldons, panchenko1997foldon}. 

Exons have also been compared with secondary structure elements, with negative results \cite{stoltzfus1994testing,weber1994intron}. Evidence of a co-occurrence between exon boundaries and protein domain border positions has been found by others and used to support the exon-shuffling theory \cite{liu2004protein,smithers2019genes}. At least for some genomes, it has been shown that this correlation of domains and exons can not be explained with a neutral null model \cite{cui2021evidence}.

The search for folding elemental units on some particular proteins has been pursued directly with various experimental methods and models. Foldons have been identified for Cytochrome C through Hydrogen exchange experiments \cite{weinkam2005funneled,maity2005protein}. These agreed with those found computationally with a perfectly funneled energy model \cite{weinkam2010folding}. Dihydrofolate reductase (DHFR) has been analyzed by molecular dissection \cite{gegg1997probing}, circular permutation  \cite{iwakura2000systematic}, systematic Alanine insertion \cite{shiba2011systematic} and overlapped contact volume  \cite{takase2021structure}, leading to potential modular decompositions.

Folding units are not necessarily continuous in sequence. Secondary structure motifs have been grouped into overlapping foldons \cite{lindberg2007malleability} and physically connected amino acids in the tertiary structure have been correlated into 'protein sectors' \cite{halabi2009protein}.

In the case of repeat-proteins, their structural symmetry allows a way to naturally define folding units for an entire protein family \cite{schafer2012discrete, ferreiro2008energy}. Remarkably, by modeling the interactions between these minimal common foldons, different groups of elements that fold at the same time emerge naturally for each protein, defining domains that coincide with those described experimentally \cite{galpern2022evolution}. Interestingly, it has been seen that repeat-proteins are made of exons that encode one or two complete repeats, exhibiting a striking conservation of intron position and phase \cite{street2006role}. 

In this work, we revisit the concept of exon regions as potential protein folding modules. 
By leveraging gene annotation and protein sequence databases, we explore exon conservation across 38 protein families to assess whether exons exhibit foldon-like behavior through energy landscape measures. To accomplish this, we use the coarse-grained forcefield AWSEM \cite{davtyan2012awsem} to establish a quantitative score that assesses the independence of foldability for sequence fragments. Furthermore, we investigate a systematic partitioning of proteins into non-overlapping units using exon boundary hotspots.

\section*{Results}
\subsection*{Exons as protein segments}

We mapped exon positions to the amino acid sequence in the multiple sequence alignments (MSA) for 38 protein domain families. Details about the data for each family are summarized in Table \ref{tab:protein_info}, with curation specifics available in Methods. We divided the protein sequences into the segments that are encoded by each exon. It is noteworthy that the distribution of exons per protein in this set follows an exponential pattern (see Fig \ref{fig:fig1s}). The distribution of exon sizes for the entire set also exhibits an exponential decay, a result expected under the assumption that intron positions are the result of independent trials of a neutral stochastic process (see Fig \ref{fig:fig1s}).

\begin{figure*}[h!]
\centering
\includegraphics[width=14.6cm]{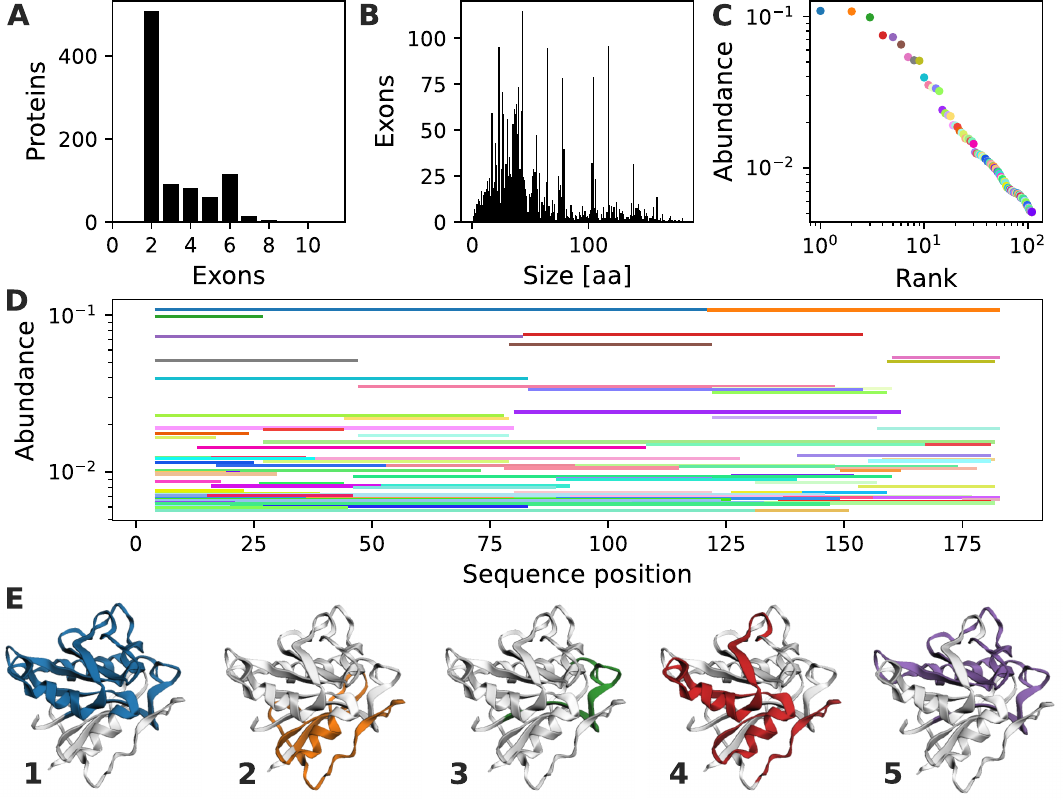}
\caption{\textbf{Exon characterization for DHFR family}. A. Number of exons per protein. B. Exon size distribution, measured in the amino acids of the corresponding protein segment. C. Abundance-rank plot, including exons present in at least 0.5\% of the effective sequences. D. Abundance as a function of the aligned sequence position for the exons in panel C. E. Projection on the 3D family reference structure (PDB: 8dfr) for the most abundant exon (blue), the second one (orange), the third one (green), the fourth one (red) and fifth one (purple). Color assignment to exon is shared between panels C, D and E.}
\label{fig:fig1}
\end{figure*}

However, when we focus our analysis on individual protein families, we observe deviations from the general trends. We present specific results for the DHFR family in Figure \ref{fig:fig1} for illustrating this phenomenon. In the DHFR family, certain exceptionally large exon sizes are overrepresented (Panel B), suggesting that natural selection may influence exon lengths. Results for other families are presented in Fig \ref{fig:fig3s}. Interestingly, none of the families individually shows a clear exponential decay trend. Instead, some preferred exon sizes stand out. For structurally symmetric domains like the Cristall or MCH-I family, a characteristic exon length emerges and the exponential decay is not present at all.

Along the MSA, exon positions are sometimes exactly conserved allowing us to measure exon relative abundance. Abundance-rank plots present power-law trends, which can be a consequence of spreading phylogenetic diversity represented by exons. The DHFR case is shown in Figure \ref{fig:fig1}C-E.  We see that the two most abundant exons (blue and orange) are present in 10\% of the sequences, causing a division of the domain at residue 119 into two consecutive fragments. The fourth (red) and the fifth (purple) most frequent exons define an almost completely alternative partition of the structure. In contrast, some other exons while abundant do not always come along with a specific exon in the complementary part of the chain. An example of this is the third most abundant exon (green). This pattern suggest the existence of multiple alternative options that can complete the open reading frame. 

\subsection*{Exon foldability}
\begin{figure*}[h!]
\centering
\includegraphics[width=7cm]{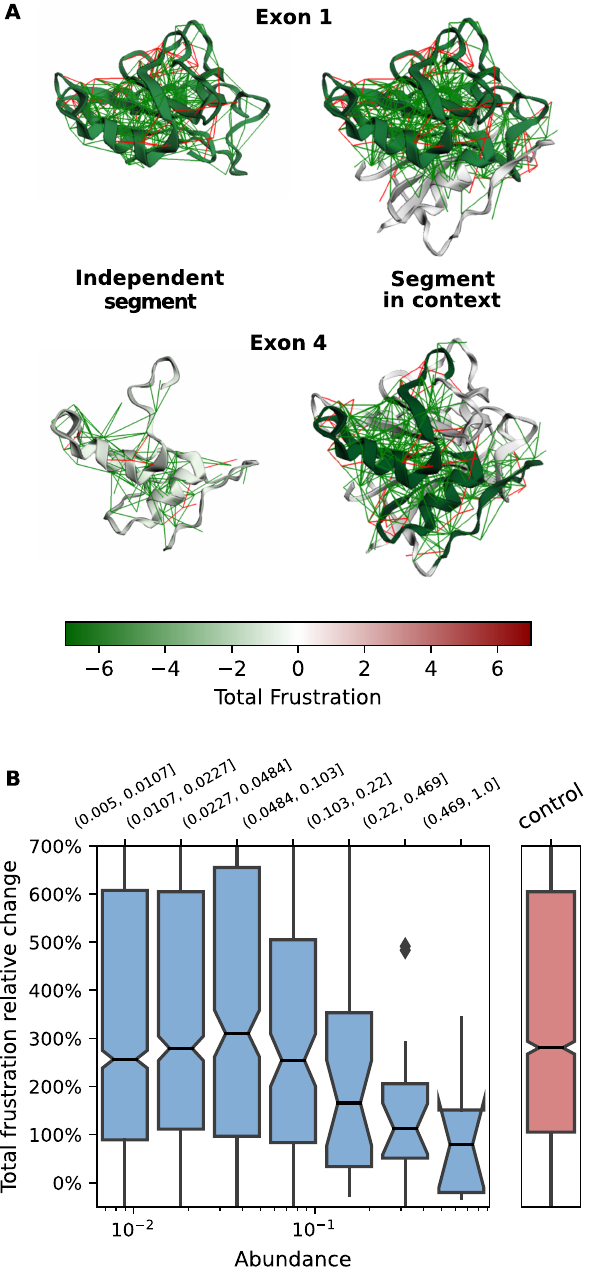}
\caption{\textbf{Relative change in total frustration.} A. Definition, using as examples DHFR exons 1 and 4 (blue and red in Fig \ref{fig:fig1}E). The relative change in total frustration of a protein segment $\delta f$ compares it in two schemes, as independent segment (left) and the same segment in the context of the whole protein (right). We color the segments in the reference structure (PDB: 8dfr) according to the total frustration. For exon 1, the segment in context ($f_{C1}=-5.47$) is as frustrated as the independent scheme ($ f_{I1}=-5.05$), the relative change $\delta f_1 = 8\% $ is small, the exon as minimally frustrated as it can be without the rest of the protein. But for exon 4, $f_{C4}=-6.59$ while $f_{I4}=-0.43$, therefore the relative change $\delta f_4 = 1446\%$ is huge. Contact frustration is added for indicating the contacts that stabilize (green) or destabilize (red) the segment in each scheme. B. Total frustration relative change as a function of exon abundance for all the families together, in a box plot in logarithmic scale. Blue boxes contain the central 50\% of data, with a black line in the median and a notch indicating its confidence interval. Abundance interval for each box is indicated on top. The red box on the right represent the distribution of a control group of alternative exons, sampled from each family size distribution. Below an abundance of 5\%, natural exons distribution is indistinguishable from the control one. Over that frequency, the frustration relative change smoothly decrease.}
\label{fig:fig2}
\end{figure*}

Do natural exons behave as foldons? Foldons have been defined as quasi-independent foldable protein segments \cite{panchenko1996foldons}. A foldon then should be at least as minimally frustrated by itself as in the context of the whole protein that contains it. We therefore examine exons comparing their frustration using two schemes (Fig \ref{fig:fig2}A). In one scheme, the protein segment encoded in an exon is treated as a totally independent polymer folding to its final three-dimensional structure. In the other scenario, the folding of the same segment is treated in the context, still interacting with the rest of the protein that contains it. Using both the independent  scheme (I) and the context scheme (C) we compute the correspondent total frustration index, a Z score defined as $f = \Delta E / \delta E $, where $\Delta E$ is the energy gap between the native configuration and the molten globule state, represented by a set of decoys, and $\delta E^2$ is the energy variance of those decoys \cite{ferreiro2007localizing}. The quantities are related to the characteristic transition temperatures of the chain segments through the configurational entropy loss upon folding from a compact molten globule $S$. For the protein to be foldable on a relevant timescale, the folding temperature ($T_f \propto \Delta E / S$) should exceed its glass transition temperature ($T_g \propto \delta E / S^{1/2}$). Protein foldability, which has been used to search foldons \cite{panchenko1996foldons,panchenko1997foldon}, can be written as $\Theta = f S^{1/2} \propto T_f / T_g$.

Here, we have employed the mutational frustration index, where decoys are scrambled versions of the original sequence. The energies are computed using the coarse-grained AWSEM potential \cite{davtyan2012awsem}. The exon energy is averaged over all the sequences in the alignment that share that same exon. A single PDB structure is used as reference for each family, threading the corresponding sequence each time. We take the independent segment to retain the structure that it has in context. Details of the implementation are provided in Methods. 

We introduce $\delta f$, the relative change in total frustration of a protein segment in the transition from the independent (I) to the in context (C) scheme
\begin{equation} 
    \delta f = \frac{f_C - f_I}{-|f_I|}.  
\end{equation}
If the configurational entropy loss $S$ is the same in the two scenarios, the relative change in the total frustration can be seen also as the relative change in the foldability $\Theta$ and in the ratio $T_f/T_g$.

In Fig \ref{fig:fig2}A we present two examples. One one hand, DHFR exon 1 -the most conserved exon in Fig \ref{fig:fig1}E1- shows a small change in total frustration $\delta f_1 = 8\% $. It's a segment that in isolation is almost as minimally frustrated in the context of the whole structure, as the foldon definition requires. On the other hand, exon 4 -which is slightly shorter and less conserved than exon 1- has a huge relative change $\delta f_4 = 1446\%$. In this case, the exon present in the context numerous stabilizing contacts between the fragment and the rest of the protein that minimize the total frustration. Those stabilizing native interactions are absent for the independent segment. By itself, the segment is notably less foldable, making it difficult to characterize it as a foldon. 

To see if there is a systematic relationship between the foldability independence and the exon conservation, for each protein family we compute $\delta f$ for all the exons having an abundance greater than 0.5\% and for a control group made of exon alternatives sampled from the family size distribution. Considering all families together, the relative change in total frustration median decreases with exon abundance, as Fig \ref{fig:fig2}B box plot shows. Below an abundance of 5\%, $\delta f$ distribution for natural exons is not distinguishable from the control group distribution. But for the more abundant exons, the frustration relative change starts a descending trend. This effect does not directly result from exon length, which does not significantly change with abundance (Fig \ref{fig:fig4s}). Most conserved exons are more likely to behave as foldons than do the less abundant exons.

\subsection*{Minimal common exons}

\begin{figure*}[h!]
\centering
\includegraphics[width=14.6cm]{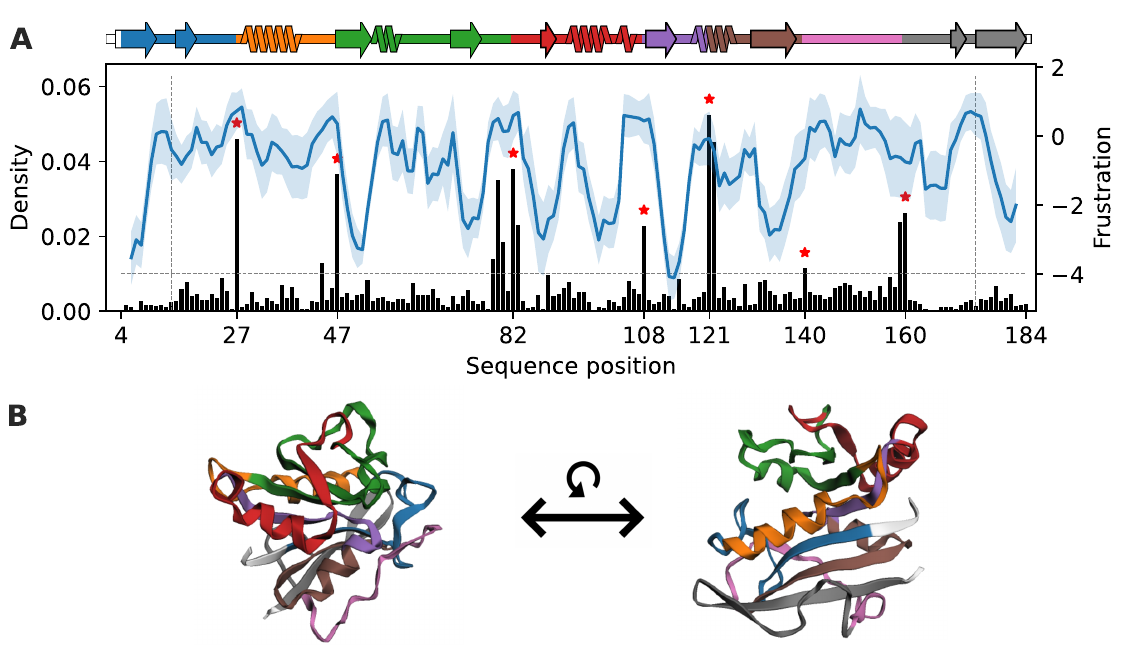}
\caption{\textbf{Exon boundary analysis for DHFR}. A. Histogram of exon boundaries (black bars). Boundary hotspots,  histogram local maxima, are marked with red stars. Below the horizontal dashed grey (density = 0.01) line we ignore the peaks, considering them background noise. We also ignore peaks closer than 10 residues from each other or to alignment limits (vertical dashed grey lines). Over the histogram, the local smoothed frustration signal (blue), its average over the sequences as a solid line and standard deviation as a shadow. On top, the secondary structure representation of the reference structure of the family (PDB: 8dfr). Colors represent the minimal common exons (MCE), the sequence partition given by boundary histogram hotspots. Almost all the MCE are made of uninterrupted secondary structure elements or combinations of them B. Minimal common exons projected on the reference 3D structure with the same colors used in panel A, with two different orientations of the structure.}
\label{fig:fig3}
\end{figure*}

Exon boundaries are not evenly distributed along the sequences. We present a histogram of exon boundary positions for the DHFR family as a case study in Fig \ref{fig:fig3}A (black bars). We note that there are no absolutely prohibited positions for the exon boundaries when one considers the entire sequence alignment. In addition, high frequency hotspots appear every 20 to 40 residues. Taking into account the exon size distribution for DHFR (Fig \ref{fig:fig1}B), the hotspots are too close to each other to be explained just by repeating some very abundant exons. Instead, they reveal an overlap of different sequence partitions. The hotspots can be interpreted as alternative breakup points in the exon-intron structure, conserved through the family. A similar pattern is seen for the other studied families (see Fig \ref{fig:fig6s} and \ref{fig:fig7s}).

The local maxima in the histogram of Fig \ref{fig:fig3}A (red stars) can be used to divide each protein domain (and the corresponding MSA) into a set of segments, that we call \emph{minimal common exons} (MCE). We identify the minimal common exons with different colors along the secondary structure description on top the histogram of Fig \ref{fig:fig3}A and the PDB structure in Fig \ref{fig:fig3}B.

The relationship between the MCE and secondary structure stands out in this case. With a single exception (position 121) the hotspots do not break alpha helices or beta strands, instead the breaks occur in coil-like regions. Equivalently, one can describe the MCE as being complete secondary structure elements or combinations of them.  

We compare this picture with a neutral model, where alternative exons are generated by sampling the exponential size distribution of MCE from each and every family (see Methods). A Z score comparing the natural MCE and those alternative pieces reveals that for the majority of the families that we studied (including DHFR) the actual boundaries occur more than expected in coil-like regions and rarely occur in alpha or beta elements (Fig \ref{fig:fig4}). 

A local smoothed frustration signal can be defined computing the total mutational frustration for a segment of 5 residues on a sliding window (Fig \ref{fig:fig3}A, blue line). This signal shows some correlation with secondary structure categories. The beta regions generally have lower frustration than the rest of the structure on average (Fig \ref{fig:fig5s}). A Z score comparing natural MCE and pieces sampled from a neutral model shows that frustration is higher than expected on boundaries for the majority of families (Fig \ref{fig:fig4}), but there are some exceptions to the pattern.

We compute $\delta f$, the relative change in total frustration for the MCE. A comparison of the MCE $\delta f$ distribution with that generated by the neutral model yields heterogeneous results. Only around one third of the families that we studied have significantly more independently foldable MCE than the neutral model would give (Fig \ref{fig:fig4}).

We see that MCE are not as independently foldable as actual exons. They seem to be too short to be independent from the rest of the protein. Instead, they work as fundamental units that can alternatively combine into different bigger segments (the real exons). Within these possibilities, the most frequent ones stand out as more independently foldable that random segments.

\subsection*{Family specific characteristics}
\begin{figure*}[h!]
\centering
\includegraphics[width=7cm]{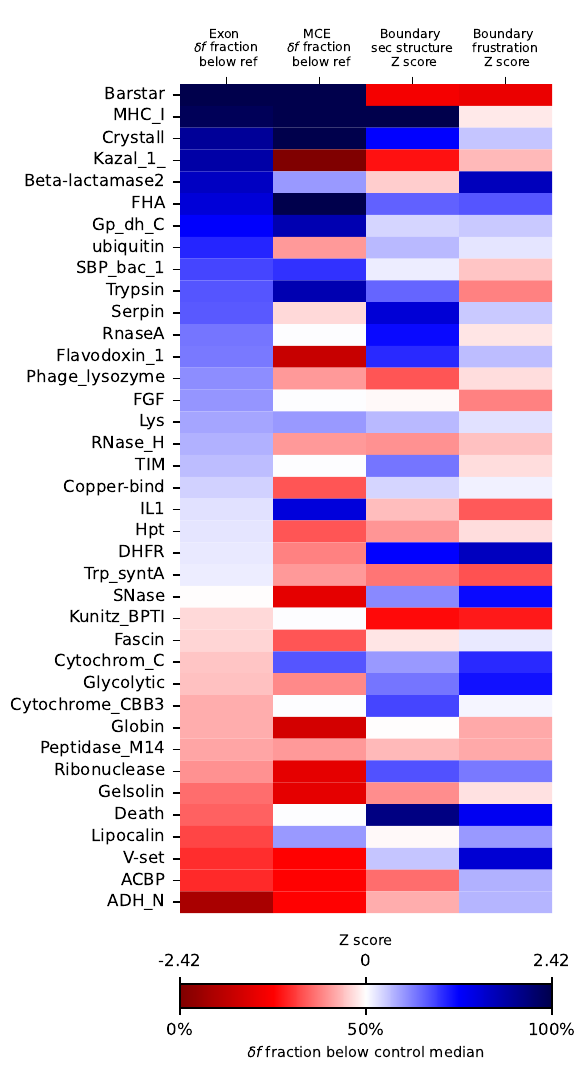}
\caption{\textbf{Summary for each family}. We include two groups of results in heat maps, each one with the corresponding scale at the bottom. First and second columns on the left contain represent the fraction of $\delta f$ distribution below the family control group median for the actual exons (first) and the minimal common exons (second). This scores go from 0\% (red), where all the exons are less independently foldable than the reference, to 100\% (blue) where all the exons are more independently foldable than it. The families are sorted using the first column score. The last two columns on the right represent Z scores comparing boundary hot spot positions (MCE boundaries) with alternative ones generated with a neutral model. In the third column, the score is positive (blue) when boundaries occur more than expected in coil-like regions. In the last column, the score is positive (blue) when boundaries occur in regions that are more frustrated than expected. }
\label{fig:fig4}
\end{figure*}

\begin{figure*}[h!]
\centering
\includegraphics[width=14.6cm]{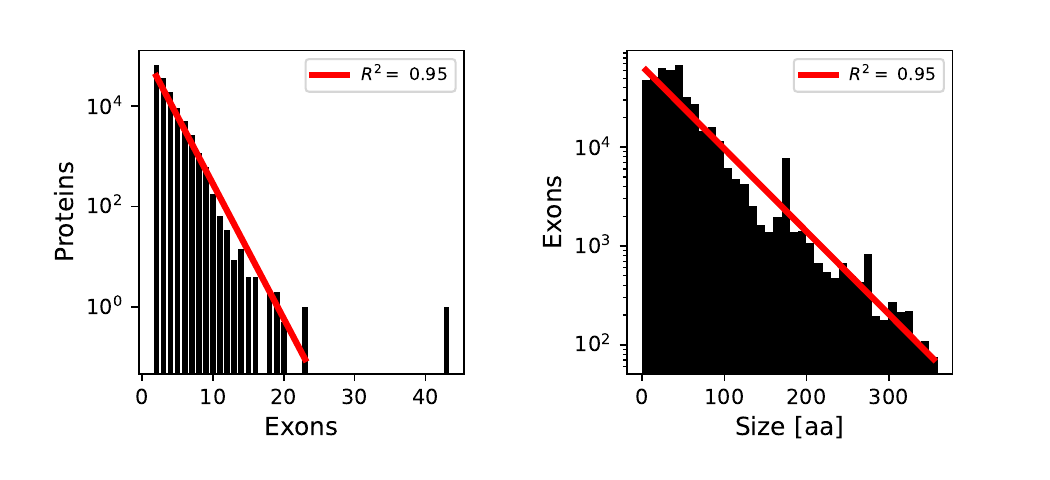}
\caption{\textbf{Exon characterization, all the families together.} A. Number of exons per protein. An exponential fit is shown in red. B. Exon size distribution, measured in the amino acids of the corresponding protein segment. An exponential fit is shown in red.}
\label{fig:fig1s}
\end{figure*}

We summarize the results obtained from four different approaches for the actual exons of the 38 studied families in a heat map (Fig \ref{fig:fig4}). 
The first heat map column on the left represents the fraction of the $\delta f$ distribution having values below the median of the alternative exons control group, which is family-specific. This score runs from 0\% (red), where all the exons are less independently foldable than the alternative exon sampled reference, to 100\% (blue) where all the exons are more independently foldable than in the reference. Families are sorted according to this statistic. For the majority of the families studied the natural exons are more independently foldable than would be expected. But the results are diverse and for some families the alternative pieces are more independently foldable than the natural ones. 

We compare the $\delta f$ for the minimal common exons (MCE) with the correspondent control group. The fraction of the $\delta f$ distribution below the control group median for MCE is shown in the second column of the heat map (Fig \ref{fig:fig4}). With a few exceptions (IL1, Cytochrom C, Lipocalin), the actual exons show higher scores than do the MCE of the same family.  

The third column represents the Z score that compares MCE boundaries to those of a neutral control group. A positive Z score (blue) indicates there are more boundaries in coil-like regions than what would be expected based on a neutral model, while a negative score (red) would indicate that there are more boundaries in the stable secondary structure regions. Finally, the last column shows whether the MCE boundaries are more frustrated than expected, measuring frustration using a 5-residues sliding window. 

We find that protein families display several patterns when we take together the analysis of exon foldability and boundary occurrence regions. For some families, MCH I, Crystall, Beta-lactamase, FHA, Gpdh-C and Trypsin, exons are independently foldable and codify mostly uninterrupted stable secondary structure regions. In another set of families the most common boundaries are clearly not random, but exon folding does not seem to be the most relevant signature for their evolutionary selection. This is the case for Ribonuclease, Death, V-set, Glycolytic and SNase.  There are some other examples however, Barstar, IL1, Kazal and Phage lysozyme, where the actual exons are more independently foldable than expected, but their boundaries do not particularly occur in highly frustrated or coil-like regions.

In the case study presented previously in this work, DHFR, as in ubiquitin, Serpin and Flavodoxin, we find that the actual exons are more independently foldable than expected, but this is not the case for the minimal common exons. Interestingly, these minimal common exons are mostly contiguous secondary structure elements that may not fold by themselves but can combine into larger segments -the actual exons- that are less frustrated.
  
\section*{Concluding Remarks}
We have revisited the correspondence between exons and protein folding modules. By mapping the exon-intron boundaries to multiple sequence alignments, we identified conserved exon partitions. For each protein family, the size distribution of exons deviates from exponential decay due to particular and very common instances. A neutral model, where intron positions are chosen through sequential independent trials of a stochastic process, cannot explain these patterns. Through frustration analysis, we found that protein segments corresponding to the most common exons are clearly more independently foldable than others. On average, the size of the foldable fragments does not change with exon abundance. Presumably, natural selection acting on exons is influencing the size distribution by taking into account the folding of the corresponding protein fragment. If exons have been shuffled during evolution, the foldability independence of the protein region encoded by an exon becomes an advantageous feature, allowing it to be inserted in a different topology or copied in tandem and maximizing the chances of giving rise to a foldable polymer.

The most common exons can function as folding units. Nevertheless, it's important to note that these exons may not always span the entire protein domain; instances exist where they overlap with each other. We define a systematic way of partitioning a multiple sequence alignment into non-overlapping segments using the exon boundary histogram hot spot positions along the sequence. These selected hot spots divide the protein into minimal common exons (MCE). For the majority of the studied families, the MCE consists of uninterrupted alpha and/or beta elements, and the boundaries between them occur in highly frustrated or coil regions. This co-occurrence has been previously studied in earlier works, but a no significant tendency to co-occurr was reported \cite{stoltzfus1994testing,weber1994intron}.

While it has been observed that domain boundaries may match exon boundaries \cite{liu2004protein,smithers2019genes,cui2021evidence}, our results show that there is an internal structure within the protein domains. The most conserved intron positions define possible splitting points for the actual modules of a protein domain. The diversity of exons within a protein family arises from the alternative usage of these breaking points, forming the actual exons. It should be noted that each family may have a different evolutionary history, where the exon boundaries may be seen as scars of that history and may be maximizing the chances of giving rise to a new fold. We have shown that in certain families, conserved exon boundaries clearly delineate secondary structure elements, whereas in other families, exon frustration is remarkably minimal. We propose that both aspects must be taken into account when analyzing the relations between protein folding and evolution of particular protein domains.

\section*{Methods}

\subsection*{Data curation}
Protein multiple sequence alignments (MSA) for each family were obtained in December 2022 from Pfam \cite{pfam2015}, now hosted by INTERPRO database \cite{interpro}. For minimizing phylogenetic bias within each MSA, we clustered by full sequence similarity using CD-hit \cite{cdhit} at 90\% cutoff and we assigned a weight to each sequence defined as $1/n_i$, being $n_i$ the number of sequences in the $i$th cluster. All the statistics were made taking into account these sequence weights.
We used a target 3D structure selected from the PDB \cite{berman2000protein} for each family (see Table \ref{tab:protein_info}) and we aligned the MSA to its corresponding sequence, keeping only the positions of the MSA that are present in the target sequence. To summarize, MSA positions are Pfam domain positions in the target PDB structure. All the calculations that involve the protein tertiary structure were made using the target PDB structure selected for the family. Secondary structure data were obtained using the DSSP algorithm \cite{dssp} on the target structures. 
Exon data were obtained from GenBank database \cite{genbank}. We downloaded all the gene files corresponding to the Uniprot IDs \cite{boutet2007uniprotkb} in our MSAs, excluding Bacteria. Single-exon sequences were excluded from the analysis. We parsed the gene files to get the exon positions and we mapped them to the corresponding MSA, obtaining the amino-acid sequence segment corresponding to each exon. Every exon starting and ending position was referenced to the MSA. We calculated the exon relative abundance as the sum of the sequence weights of all the sequences that have an exon in the same position of the MSA, normalized by the sum of the sequence weights of the protein family.    
Data download and curation were carried out using python scripts. The code is available at GitHub: \url{https://github.com/eagalpern/exon-foldon}.

\subsection*{Total frustration relative change}
To determine the energy of a protein segment we used the AWSEM coarse-grained forcefield, including only the burial and the contact terms \cite{davtyan2012awsem}. We used a single 3D target structure for each family and we threaded it with the particular sequence we wanted to evaluate. The total energy of a segment is calculated according to two different scenarios. The independent (I) scheme energy includes the contact terms of all the pairs within the segment, while the in-context scheme (C) considers also all the contacts between segment residues and other protein positions outside it,
\begin{subequations}
\label{eq:awsem_ie}
\begin{align}
    H_{I} =\sum_{i=a}^{b}H_i^{burial} + \sum_{i=a}^{b} \sum_{j=a}^{b}H_{ij}^{contact}  \\
    H_{C} =\sum_{i=a}^{b}H_i^{burial}  + \sum_{i=a}^{b} \sum_{j=1}^{L}H_{ij}^{contact},
\end{align}
\end{subequations}
where the segment goes from position $a$ to $b$, within a sequence of $L$ residues. For each exon, segment energy is a weighted average over all the sequences in the alignment that have the exon. We exclude from the average exons where gaps represent more than 50\% of their sequence. The frustration $f$ was calculated using decoy sets, constructed by randomizing the identity of the amino acids of the complete sequences. Decoy sequences were also threaded through the same tertiary structure selected for the family.
Energy calculations were made using a python implementation of the protein frustratometer \cite{rausch2021frustratometer}, available at Github: \url{https://github.com/HanaJaafari/DCA_Frustratometer}. Total frustration calculation scripts are available at GitHub: \url{https://github.com/eagalpern/exon-foldon}.

\subsection*{Local frustration}
The local frustration of position $x$ was calculated evaluating the total frustration (as defined previously) of a 5-residues segment centered on $x$. The signal was obtained by sliding this 5-residues window along the sequence. We consider each segment in the context of the whole protein. 

\subsection*{Boundary local maxima searching criteria}
We searched for relative maxima in the exon boundary histograms comparing positions with other 10 to each side. We discarded any maximum closer than 10 positions to the beginning or to the end of the sequence, and also positions with histogram density below 0.01.

\subsection*{Visualization tools}
Secondary structure linear visualizations were made adapting SSDraw python library \cite{chen2023ssdraw}.
Tertiary structure visualizations were made using py3Dmol python library \cite{rego20153dmol}.

\subsection*{Exon control groups}
For each family we made a size-wise control group for exons, a set of segments whose size was obtained by sampling the family's exon size distribution along with a random initial position. For each of these segments, or exon alternatives, we randomly assigned 100 natural sequences from the alignment that can be threaded along the reference tertiary structure to compute energy, frustration and frustration relative change. For the minimal common exons (MCE) of each family, another set of exon alternatives were defined. Given that there are only a few MCEs per family, MCE alternatives are obtained generating consecutive segments sampling its sizes from a geometrical distribution fitted from all families MCE sizes. As MCE are larger than the minimum distance we imposed between histogram maxima (10 residues, see above), we eliminate MCE alternatives shorter than 10 residues.

\subsection*{Fraction scores}
The distribution of the total frustration relative change for exons $\delta f_ {exon}$ was compared for each family with the corresponding size-wise control group distribution. We used as a score the weighted fraction of $\delta f_ {exon}$ below the median of the distribution $\delta f_{control}$, given by
\begin{equation}
  frac.score^{exon} =  \sum_{i} \, w_{i} \, \delta_{i, \, \delta f_{i} < median(\delta f_{control})},
\end{equation}
where $w_i$ is the abundance of the exon $i$ and $\delta_{i,x}$ is the Kronecker symbol, taking value one if the condition $x$ is True for the exon $i$ and zero otherwise.

For the minimal common exons (MCE), the reference is given by the median of $\delta f_{control}$ for the MCE control group
\begin{equation}
    frac.score^{MCE} = \sum_{i} \delta_{i, \, \delta f_{i} < median(\delta f_{control})} / N,
\end{equation}
where $i$ are the MCE and $N$ is the number of MCE for the family.

\newcommand{\<}{\langle}
\renewcommand{\>}{\rangle}

\subsection*{Boundary secondary structure Z score}
The occurrence of exon boundary local maxima (or minimal common exon boundaries, MCE) on coil-like regions (not alpha or beta) for a family was compared to the occurrence on the alternative partitions that define the MCE control using a Z score defined as
\begin{equation}
Zscore ^{coil}=  \frac{ {\bar{\delta}_{MCE}}  -\<\bar{\delta}_{control} \> }{\sigma_{\delta_{control}}},
\end{equation}
where $\delta$ is the Kronecker symbol, $\bar{*} $ represents the average over the partition, $\< * \> $  the average over all the decoy partitions  and $\sigma$ the standard deviation. We consider the boundary as being the ending position of each segment (MCE or control) and the first position of the next one. If at least one of them is not an alpha or beta region, we take that boundary $i$ as a positive case $\delta^i=1$, while if the two of them are beta and/or alpha, $\delta^i=0$.

\subsection*{Boundary frustration Z score}

Local frustration on exon boundary local maxima (or minimal common exon boundaries, MCE) for a family was compared with the frustration on the alternative partitions that define the MCE control  using a Z score defined as
\begin{equation}
Zscore ^{f}=  \frac{ {\bar{f}_{MCE}} -\<\bar{f}_{control} \> }{\sigma_{f_{control}}},
\end{equation}
where $f$ is the local frustration, $\bar{*} $ represents the average over the the boundary positions of a partition, $\< * \> $  the average over all the control partitions  and $\sigma$ the standard deviation. We consider as boundary the ending position of each segment (MCE or control) and the first position of the next one.

\section*{Code availability}
Python scripts for data download, data curation and total frustration calculations are available at GitHub:\url{https://github.com/eagalpern/exon-foldon}. 

\section*{Data availability}
All input data needed to reproduce the main results, including Fig \ref{fig:fig1}D-E and Fig \ref{fig:fig3} for the 38 protein families that we studied is available at GitHub:\url{https://github.com/eagalpern/exon-foldon}, along with a Jupyter notebook for visualization.

\section*{Acknowledgments}
This work was supported by the Consejo de Investigaciones Científicas y Técnicas (CONICET) (DUF is CONICET researchers and EAG is a postdoctoral fellow); CONICET Grant PIP2022-2024 - 11220210100704CO. Universidad de Buenos Aires UBACyT 20020220200106BA. Additional support from NAI and Grant Number 80NSSC18M0093 Proposal ENIGMA: EVOLUTION OF NANOMACHINES IN GEOSPHERES AND MICROBIAL ANCESTORS (NASA ASTROBIOLOGY INSTITUTE CYCLE 8).  P.G.W. was supported both by the Bullard–Welch Chair at Rice University, grant C-0016, and by the Center for Theoretical Biological Physics sponsored by NSF grant PHY-2019745.
\bibliographystyle{unsrt}
\bibliography{diego,bibliography,biblio_intro,byhand,repeat_prots}

\newpage
\section*{Supporting information}
\beginsupplement

\begin{table}[h]
\centering
\begin{tabular}{|c|c|c|c|c|c|c|}
\hline
\textbf{Family} & \textbf{Pfam ID} & \textbf{Interpro ID} & \textbf{PDB ID} & \textbf{UniProt ID} & \textbf{Eff. sequences} & \textbf{Eff. exons} \\
\hline
ACBP & PF00887 & IPR000582 & 2abd & P07107/3-78 & 7239 & 7778 \\
\hline
ADH\_N & PF08240 & IPR013154 & 8adh & P00327/35-160 & 177236 & 45096 \\
\hline
Barstar & PF01337 & IPR000468 & 1bta & P11540/6-81 & 4756 & 2 \\
\hline
Beta-lactamase2 & PF13354 & IPR045155 & 1g68 & P16897/41-257 & 17485 & 37 \\
\hline
Copper-bind & PF00127 & IPR000923 & 5pcy & P00299/71-168 & 13122 & 362 \\
\hline
Crystall & PF00030 & IPR001064 & 4gcr & P02526/90-171 & 10051 & 6435 \\
\hline
Cytochrom\_C & PF00034 & IPR009056 & 1cyc & P00025/4-102 & 68475 & 2208 \\
\hline
Cytochrome\_CBB3 & PF13442 & IPR009056 & 451c & P00099/25-100 & 75198 & 465 \\
\hline
DHFR & PF00186 & IPR001796 & 8dfr & P00378/4-184 & 17686 & 2779 \\
\hline
Death & PF00531 & IPR000488 & 1e3y & Q13158/101-179 & 9895 & 6209 \\
\hline
FGF & PF00167 & IPR002209 & 1rg8 & P05230/28-148 & 3432 & 3668 \\
\hline
FHA & PF00498 & IPR000253 & 1mzk & P46014/208-284 & 63595 & 18260 \\
\hline
Fascin & PF06268 & IPR022768 & 1hce & P13231/5-60 & 1722 & 1507 \\
\hline
Flavodoxin\_1 & PF00258 & IPR008254 & 5nll & P00322/3-128 & 23898 & 9560 \\
\hline
Gelsolin & PF00626 & IPR007123 & 2vil & P02640/27-108 & 18403 & 19369 \\
\hline
Globin & PF00042 & IPR000971 & 1mba & P02210/28-142 & 13154 & 7691 \\
\hline
Glycolytic & PF00274 & IPR000741 & 1ado & P00883/15-364 & 6026 & 5194 \\
\hline
Gp\_dh\_C & PF02800 & IPR020829 & 4wnc & P04406/157-314 & 23844 & 5895 \\
\hline
Hpt & PF01627 & IPR008207 & 1sr2 & P39838/811-868 & 57627 & 3198 \\
\hline
IL1 & PF00340 & IPR000975 & 2i1b & P01584/147-264 & 1144 & 1479 \\
\hline
Kazal\_1\_ & PF00050 & IPR002350 & 2ovo & P67954/8-56 & 8342 & 5956 \\
\hline
Kunitz\_BPTI & PF00014 & IPR002223 & 5pti & P00974/39-90 & 24441 & 12370 \\
\hline
Lipocalin & PF00061 & IPR000566 & 3npo & P02754/32-170 & 5755 & 9255 \\
\hline
Lys & PF00062 & IPR001916 & 2lyz & P00698/19-145 & 2035 & 1974 \\
\hline
MHC\_I & PF00129 & IPR011161 & 3hla & P04439/25-203 & 10490 & 4828 \\
\hline
Peptidase\_M14 & PF00246 & IPR000834 & 5cpa & P00730/128-405 & 36053 & 30772 \\
\hline
Phage\_lysozyme & PF00959 & IPR002196 & 1am7 & P03706/33-140 & 6558 & 485 \\
\hline
RNase\_H & PF00075 & IPR002156 & 1hrh & P03366/1037-1155 & 28278 & 5646 \\
\hline
Ribonuclease & PF00545 & IPR000026 & 1rnb & P00648/67-154 & 3999 & 657 \\
\hline
RnaseA & PF00074 & IPR023412 & 1rbb & P61823/30-147 & 1467 & 156 \\
\hline
SBP\_bac\_1 & PF01547 & IPR006059 & 1omp & P0AEX9/45-315 & 54935 & 466 \\
\hline
SNase & PF00565 & IPR016071 & 1snc & P00644/115-224 & 22705 & 7295 \\
\hline
Serpin & PF00079 & IPR023796 & 1qlp & P01009/54-415 & 19934 & 28580 \\
\hline
TIM & PF00121 & IPR000652 & 1tim & P00940/7-246 & 22860 & 6194 \\
\hline
Trp\_syntA & PF00290 & IPR002028 & 1bks & P00929/8-266 & 16260 & 2907 \\
\hline
Trypsin & PF00089 & IPR001254 & 3ptn & P00760/24-239 & 83202 & 111447 \\
\hline
V-set & PF07686 & IPR013106 & 1rei & P01593/28-111 & 85406 & 31509 \\
\hline
ubiquitin & PF00240 & IPR000626 & 1ubq & P0CG48/611-682 & 28986 & 24488 \\
\hline
\end{tabular}
\caption{\textbf{Summary of Protein Domain Families.} Information on 38 protein domain families, including their Pfam family name and ID, Interpro ID, PDB ID, UniProt ID, the number of effective sequences, and the effective number of exons. Details about data curation are available in the Methods section}
\label{tab:protein_info}
\end{table}

\begin{figure*}[h!]
\centering
\includegraphics[width=14.6cm]{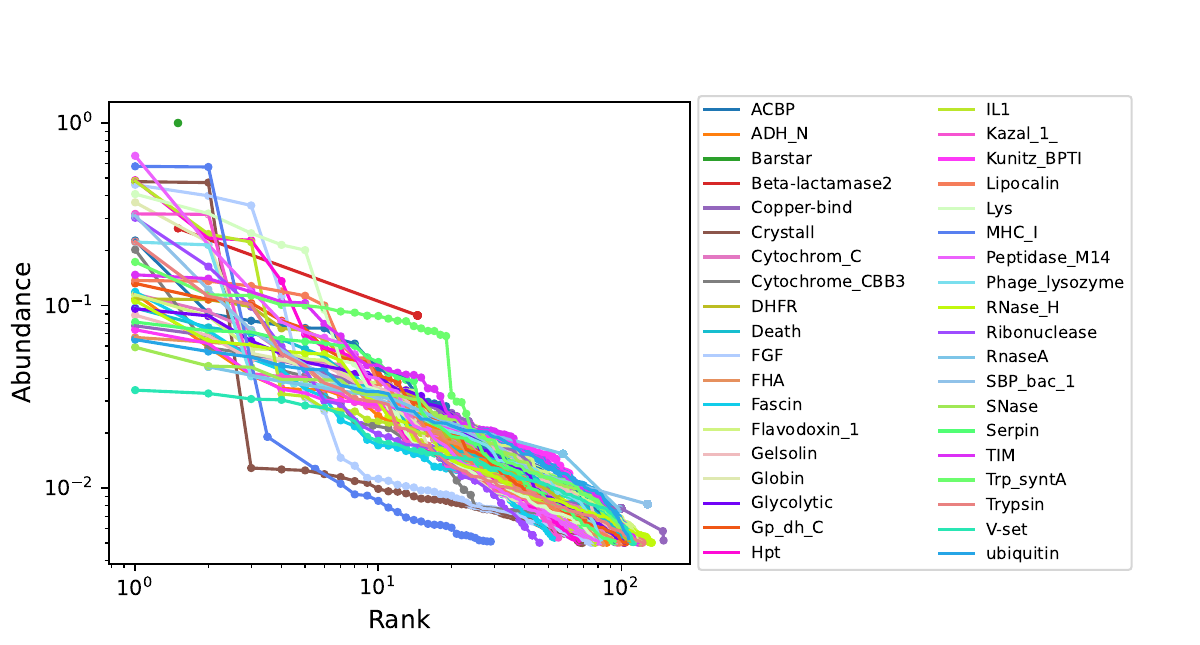}
\caption{\textbf{Abundance-rank plot for the 38 protein families that we studied.} Most of the cases present power-law relations in at least a range.} 
\label{fig:fig2s}
\end{figure*}

\begin{figure*}[h!]
\centering
\includegraphics[width=17cm]{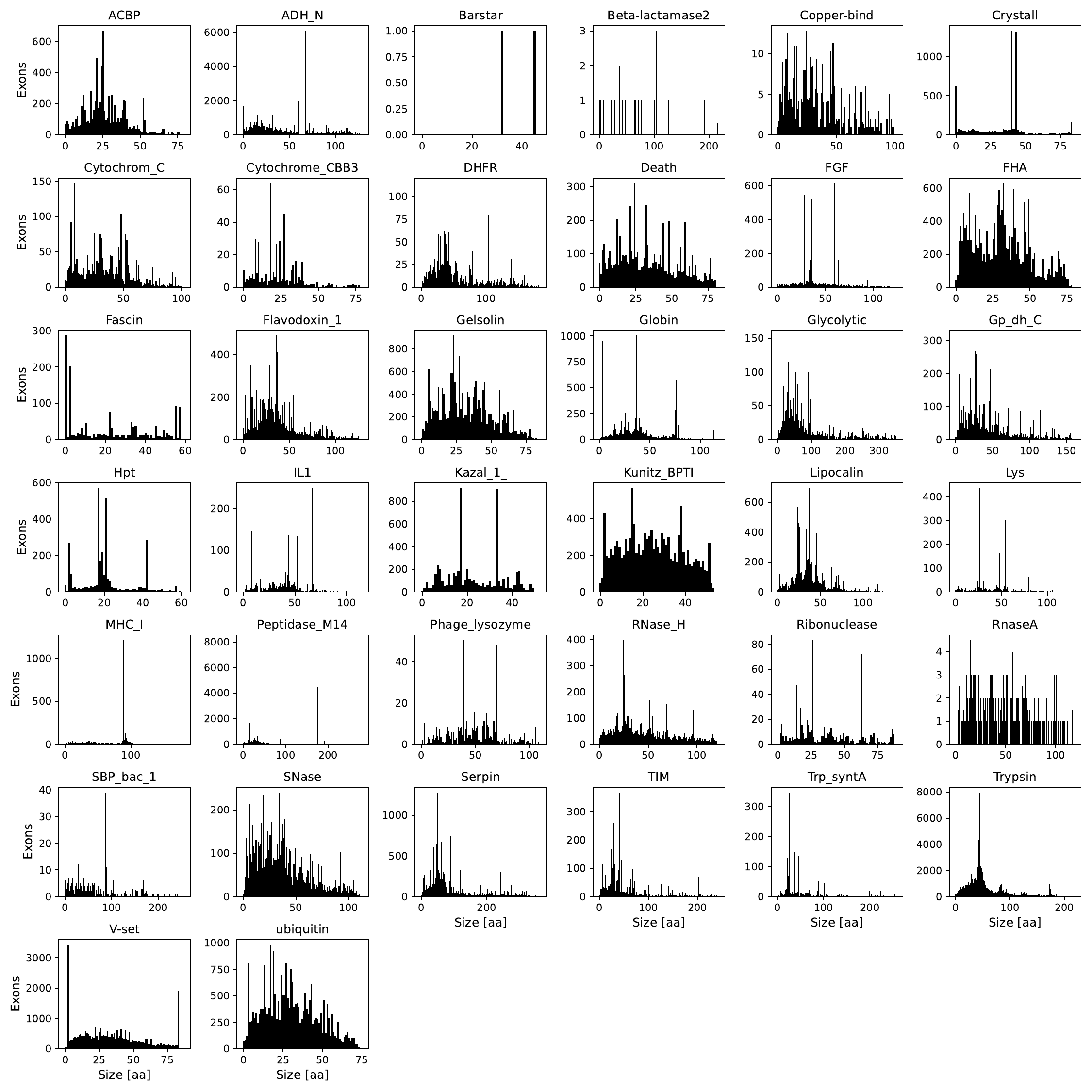}
\caption{\textbf{Exon size distribution for the 38 protein families that we studied.} Exon size is measured in the number of amino acids of the corresponding protein segment. None of the families individually shows a clear exponential decay trend. Instead, some characteristic exon sizes stand out.}
\label{fig:fig3s}
\end{figure*}

\begin{figure*}[h!]
\centering
\includegraphics[width=10cm]{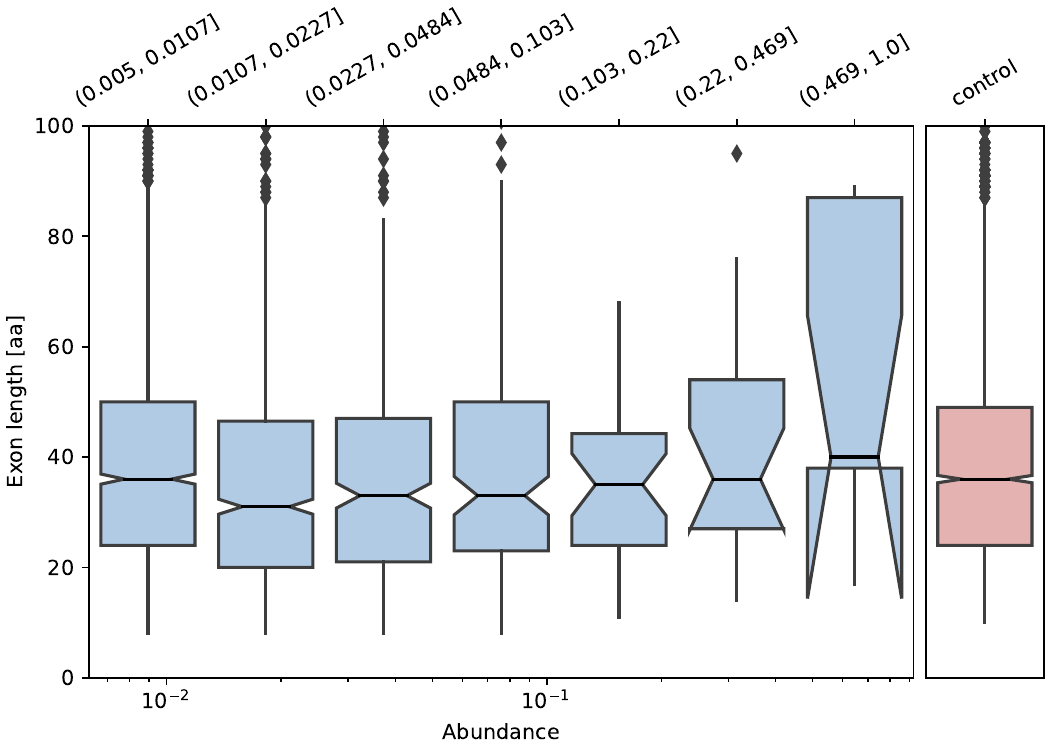}
\caption{\textbf{Exon foldability independence trend with abundance is not trivially given by exon size}. We plot Fig \ref{fig:fig3}B boxplots for exon length, in logarithmic scale. Blue boxes contain the central 50\% of data, with a black line in the median and a notch indicating its confidence interval. Abundance interval for each box is indicated on top. The red box on the right represent the distribution of exon alternatives, a control group sampled from each family size distribution.}
\label{fig:fig4s}
\end{figure*}

\begin{figure*}[h!]
\centering
\includegraphics[width=10cm]{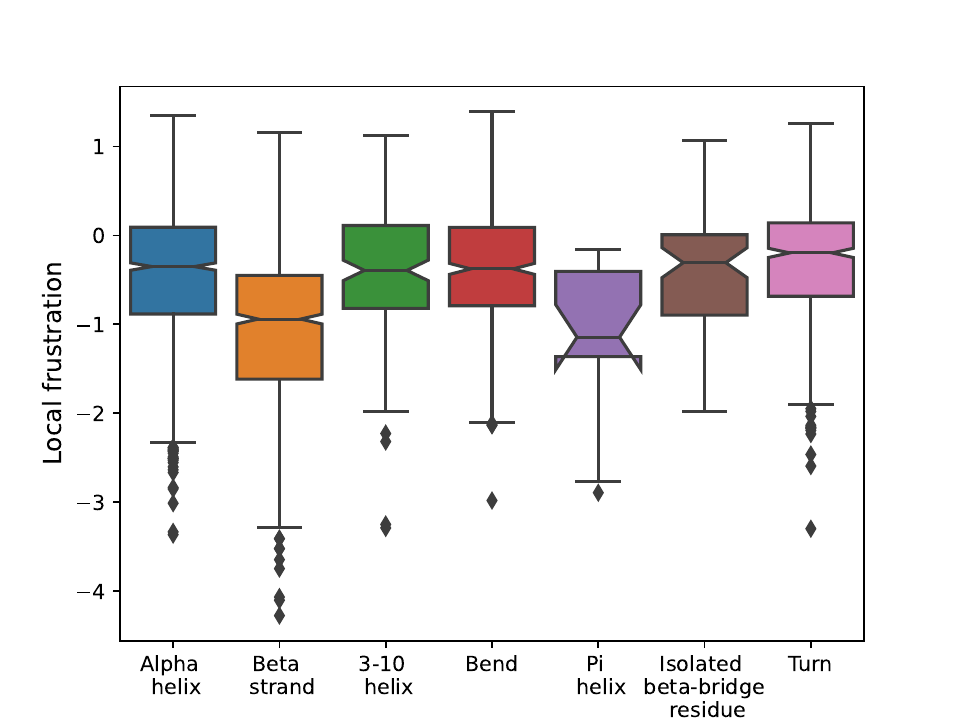}
\caption{\textbf{Local frustration and secondary structure relationship}. Local frustration is calculated on a sliding window of 5 residues, on average for all the sequences in the alignment. Secondary structure class is DSSP assignment for the center position of the window. }
\label{fig:fig5s}
\end{figure*}

\begin{figure*}[h!]
\centering
\includegraphics[width=16cm]{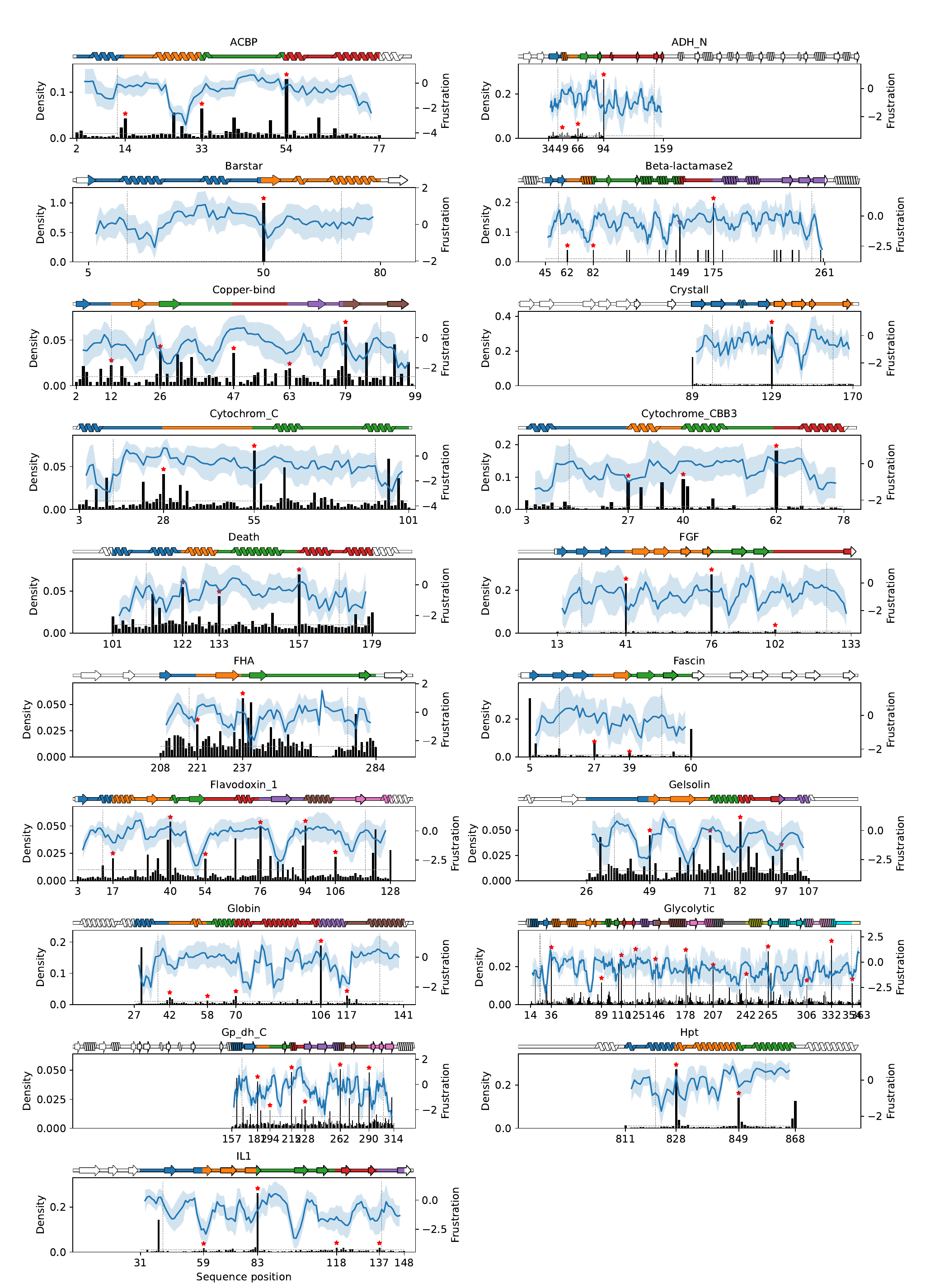}
\caption{\textbf{Exon boundary histograms, local frustration and minimal common exons (I)}. For each family, the histogram of exon boundaries (black bars). Boundary hotspots histogram local maxima are marked with red stars. Below the horizontal dashed grey (density = 0.01) line we ignore the peaks, considering them background noise. We also ignore peaks closer than 10 residues from each other or to alignment limits (vertical dashed grey lines). Over the histogram, the local smoothed frustration signal (blue), its average over the sequences as a solid line and standard deviation as a shadow. On top, the secondary structure representation of the reference structure of the family (see Table \ref{tab:protein_info} for reference PDB IDs). Colors represent the minimal common exons (MCE), the sequence partition given by boundary histogram hotspots.}
\label{fig:fig6s}
\end{figure*}

\begin{figure*}[h!]
\centering
\includegraphics[width=16cm]{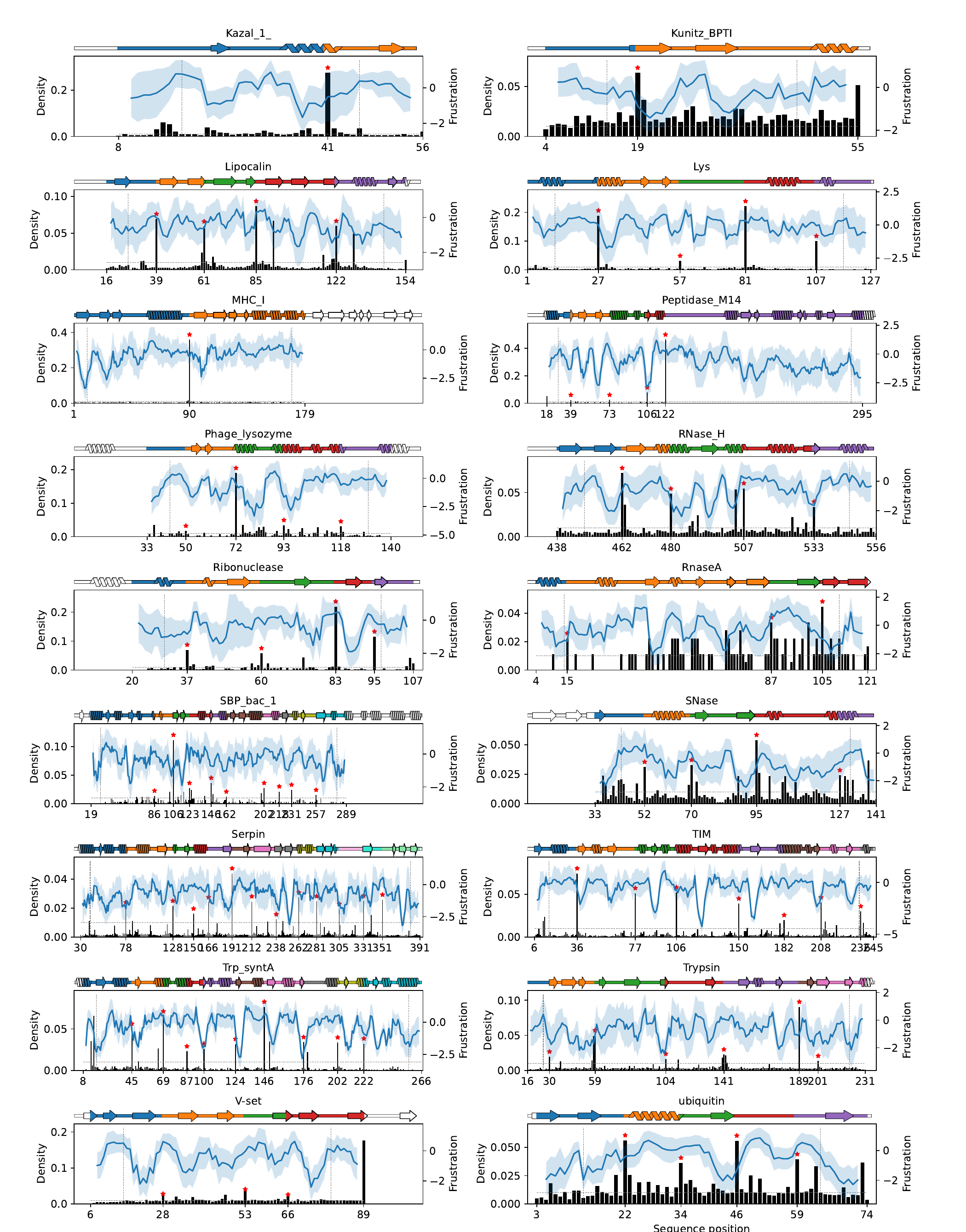}
\caption{\textbf{Exon boundary histograms, local frustration and minimal common exons (II)}. For each family, the histogram of exon boundaries (black bars). Boundary hotspots histogram local maxima are marked with red stars. Below the horizontal dashed grey (density = 0.01) line we ignore the peaks, considering them background noise. We also ignore peaks closer than 10 residues from each other or to alignment limits (vertical dashed grey lines). Over the histogram, the local smoothed frustration signal (blue), its average over the sequences as a solid line and standard deviation as a shadow. On top, the secondary structure representation of the reference structure of the family (see Table \ref{tab:protein_info} for reference PDB IDs). Colors represent the minimal common exons (MCE), the sequence partition given by boundary histogram hotspots. }
\label{fig:fig7s}
\end{figure*}

\begin{figure*}[h!]
\centering
\includegraphics[width=13cm]{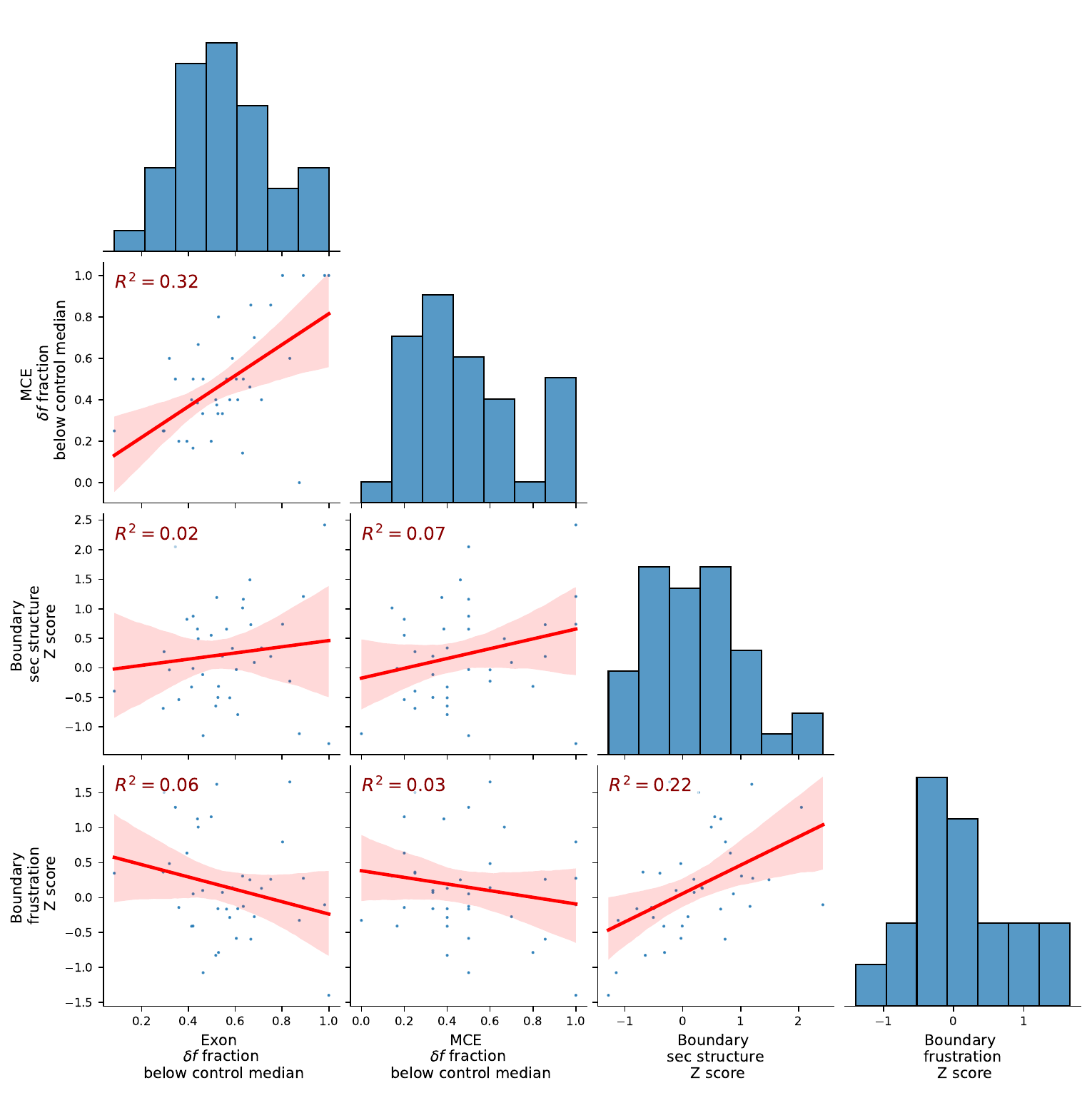}
\caption{\textbf{Scores correlation}. We include four different scores. The first two are the fraction of $\delta f$ distribution below the family control group median for the actual exons and the minimal common exons (MCE). This scores go from 0 where all the exons are less independently foldable than the reference to 1  where all the exons are more independently foldable than it. Results of these two scores for the studied families show a weak correlation ($R^{2}=0.32$). 
The third and the fourth ones are Z scores comparing boundary hot spot positions (MCE boundaries) against alternatives generated with a neutral model. The third score is positive when boundaries occur more than expected in coil-like regions, while the fourth one is positive when boundaries occur in regions that are more frustrated than expected. Results of these two scores for the studied families show also a weak correlation ($R^{2}=0.22$).}
\label{fig:fig8s}
\end{figure*}

\end{document}